\documentclass[a4paper, 12pt]{article}

\def \bea {\begin{eqnarray}}      
\def \mea {\nonumber\\}    
\def \eea {\end{eqnarray}}      

\usepackage{amsmath}     
\usepackage{epsfig,graphics}     
\usepackage{color}     
\usepackage{amssymb}     
\usepackage{latexsym}

\newcommand{\MN}{\mathcal{N}}     
\newcommand{\MW}{\mathcal{W}}     
\newcommand{\MH}{\mathcal{H}}

\newcommand{\hh}{\hat{H}}    
\newcommand{\hhaa}{\hat{H}_{\alpha\alpha}}    
\newcommand{\hhab}{\hat{H}_{\alpha\overline{\alpha}}}    
\newcommand{\hhbb}{\hat{H}_{\overline{\alpha}\overline{\alpha}}}    
\newcommand{\hhaabb}{\hat{H}_{{\alpha\alpha\overline{\alpha}}\overline{\alpha}}}    
\newcommand{\hhaaab}{\hat{H}_{\alpha\alpha\alpha\overline{\alpha}}}    
\newcommand{\hhabbb}{\hat{H}_{\alpha\overline{\alpha}\overline{\alpha}\overline{\alpha}}}    
    
\newcommand{\hr}{\hat{\rho}}

\newcommand{\hrabbb}{\hat{\rho}_{{\alpha\overline{\alpha}}\overline{\alpha}\overline{\alpha}}}    

\newcommand{\hg}{\hat{G}}    
\newcommand{\hgaa}{\hat{G}_{\alpha\alpha}}    
\newcommand{\hgab}{\hat{G}_{\alpha\overline{\alpha}}}    
\newcommand{\hgbb}{\hat{G}_{{\overline{\alpha}}\overline{\alpha}}}    
\newcommand{\hgaabb}{\hat{G}_{{\alpha\alpha\overline{\alpha}}\overline{\alpha}}}    
\newcommand{\hgaaab}{\hat{G}_{{\alpha\alpha\alpha\overline{\alpha}}}}

\newcommand{\ha}{\hat{a}}    
\newcommand{\had}{\hat{a}^\dagger}    
\newcommand{\hn}{\hat{N}}    

\newcommand{\la}{\langle}     
\newcommand{\ra}{\rangle}

\newcommand{\Tr}{\textrm{Tr}}

\begin{document}    
\begin{titlepage}     
     
\title{Semiquantum versus semiclassical mechanics for simple nonlinear systems}     
     
\author{A.J. Bracken\footnote{{\em Email:} ajb@maths.uq.edu.au} and J.G. Wood\footnote{Present address: National Centre for Immunisation Research and Surveillance, 
Children's Hospital at Westmead, Sydney 2145, Australia. {\em Email:} jamesw5@chw.edu.au}\\Department of Mathematics\\    
University of Queensland\\Brisbane 4072\\Queensland\\Australia}     
     
\date{}     
\maketitle     
     
\begin{abstract}

Quantum mechanics has been formulated in phase space, with the Wigner function as 
the representative of the quantum density operator, and 
classical mechanics has been formulated in Hilbert space, with the Groenewold operator
as the representative of the classical 
Liouville density function.   Semiclassical
approximations to the quantum evolution of the Wigner function have been  defined, enabling
the quantum evolution to be approached from a classical starting point.  Now
analogous semiquantum approximations to the classical evolution of the Groenewold operator
are defined, enabling the classical evolution to be approached from a quantum starting point. 
Simple nonlinear systems with one degree of freedom are considered, whose Hamiltonians are polynomials in the 
Hamiltonian of the simple harmonic oscillator.  The behaviour of expectation values of simple observables and of eigenvalues of the
Groenewold operator, are calculated numerically and compared for the various semiclassical and semiquantum approximations.     
    
\end{abstract}     
\end{titlepage}    
\section{Introduction}    
Developments in nanotechnology have focussed increasing attention on the interface between quantum    
mechanics and classical mechanics.    
Since the earliest days of the quantum theory, the nature of this interface has been explored by means    
of various semiclassical approximations with the introduction of quantum corrections to    
classical behaviour, characterised by expansions of quantities of interest in asymptotic    
series in increasing powers of Planck's 
constant $\hbar$\,; see for example  \cite{Wigner32,Maslov81,Voros89,Osborn95}.        
One such approach     
focusses on Wigner's quasiprobability density,    
which is the representative of    
the quantum density operator in the formulation of quantum mechanics on phase space,    
and considers quantum corrections to the classical Liouvillean evolution.       
This approach has its roots in Wigner's paper \cite{Wigner32} where    
the quasiprobability function was first    
introduced, and it     
has recently been developed    
in some theoretical detail by Osborn and Molzahn    
\cite{Osborn95},       
working in the Heisenberg picture.    
    
Recently an alternative approach has been suggested, which we refer to as    
semiquantum mechanics \cite{Bracken03}.  This     
considers the quantum-classical interface from the other side, that is to say, from the quantum side.  The    
idea is to consider a series of classical corrections to quantum mechanics, again characterised by    
increasing powers of $\hbar$.  To facilitate this, we first    
reformulate classical mechanics,     
representing classical observables on phase space by hermitian operators on Hilbert space \cite{Vercin00},     
and in particular     
representing the Liouville density     
by a quasidensity operator \cite{Groenewold46,Muga92,Bracken03} (the Groenewold operator).      
This has many of the properties of a true density operator but is not positive-definite in general    
\cite{Bracken04}.  It    
is the analogue in the Hilbert space formulation of classical mechanics, of the Wigner function in the    
phase space formulation of quantum mechanics.      
    
The evolution in time of the Groenewold operator corresponds to the classical evolution of the    
Liouville density, and     
has been given in \cite{Bracken03} as a series of terms in increasing powers of $\hbar$, with the first    
term corresponding to the familiar  quantum evolution.      
It is consideration of these terms successively that defines a series of approximations to classical    
dynamics, starting with quantum dynamics at the lowest order, and hence defining what we mean by    
semiquantum mechanics.       
    
In what follows, we shall apply these ideas to some very simple nonlinear systems in one dimension.     
These are necessarily integrable, and so incapable of showing interesting dynamical behaviour such as    
chaos.  However the nonlinearity provides a preliminary testing ground for    
semiquantum mechanics, as we consider successive classical corrections to the quantum dynamics, and    
see how the classical behaviour of expectation values of key observables emerges. We are able    
to identify some differences between semiquantum     
and semiclassical approximations  at the interface between    
quantum and classical behaviours.

\section{Semiquantum mechanics}    
    
The formulation of quantum mechanics in phase space is determined with the help of the unitary 
Weyl-Wigner    
transform ${\cal W}$, which maps each quantum observable (hermitian operator) ${\hat A}$ 
on Hilbert space into a real-valued function $A$ on phase space.
In the case of one degree of freedom, and when ${\hat A}$ is 
regarded as an integral operator with kernel $A_K(x,y)$
in the coordinate representation, the action of ${\cal W}$ 
is defined by \cite{Dubin00,Zachos02,Cassinelli03}

\bea    
A(q,\,p)={\cal W}({\hat A})(q,\,p)=\int     
A_K(q-x/2,\,q+x/2)\,e^{ipx/\hbar}\,dx\,.    
\label{wigneraction}    
\eea 
In the particular case of the density operator ${\hat \rho}$, we obtain the Wigner function
\bea
W(q,\,p)={\cal W}({\hat \rho}/2\pi\hbar)(q,\,p)\,.
\label{wigner_function}
\eea

Conversely, the formulation    
of classical mechanics in Hilbert space has been defined \cite{Vercin00,Bracken03} using the inverse    
transform ${\cal W}^{-1}$, which is    
Weyl's quantization map.  This acts on a classical observable (a real-valued    
function on phase space) $A$ to    
produce an hermitian operator ${\hat A}={\cal W}^{-1}(A)$ with kernel $A_K$ given by
\bea    
A_K(x,y)= {\cal W}^{-1}(A)_K(x,y)=    
\frac{1}{2\pi\hbar}\int    
A([x+y]/2,\,p)\,e^{ip(x-y)/\hbar}\,dp\,.    
\label{weylaction}    
\eea 
Typically ${\cal W}^{-1}$ is used only to establish the quantization of
a given classical system.  However,
we can if we wish use it to map all of classical mechanics, including classical dynamics, into
a Hilbert space formulation \cite{Vercin00,Bracken03}.  Then   
$\hbar$ in (\ref{weylaction}) can be thought of as an arbitrary    
constant with dimensions of action \cite{Muga92}, to be equated    
with Planck's constant if and when desired.

Given a Liouville probability density on phase space, that is to say a function $\rho(q,\,p)$    
satisfying    
\bea    
\int \rho(q,\,p)\,dq\,dp=1\,,\quad \rho(q,\,p)\geq 0\,,    
\label{liouville}    
\eea    
we     
define the corresponding hermitian (integral)     
operator ${\hat G}={\cal W}^{-1}(2\pi\hbar \rho)$ using 
(\ref{weylaction}).
This operator ${\hat G}$    
is the Groenewold operator, which has all the properties of a true density    
operator except that it is not in general positive-definite.     
Thus if ${\hat A}={\cal W}^{-1}(A)$, then    
\bea    
{\rm Tr}({\hat G})=1\,,    
\qquad {\rm Tr}({\hat G}\,^2)\leq 1\,,    
\qquad\qquad    
\mea    
\mea    
\langle A\rangle=\int\rho(q,\,p)\,A(q,\,p)\,dq\,dp=    
\langle {\hat A}\rangle={\rm Tr}({\hat G}{\hat A})\,,    
\label{Gproperties}    
\eea    
but not all eigenvalues of ${\hat G}$ need be positive.      
    
It was shown in \cite{Bracken03} that for suitably smooth Hamiltonians,  
the classical time evolution of the Liouville density,    
\bea    
\partial \rho/\partial t=\rho_q H_p -\rho_p H_q    
\label{poisson_evolution}    
\eea    
is mapped into the evolution    
\bea    
\frac{\partial {\hat G}(t)}{\partial t}=    
\frac{1}{i\hbar}[{\hat H},{\hat G}]    
-\frac{i\hbar}{24}\left([{\hat H}_{qq},\,{\hat G}_{pp}]    
-2[{\hat H}_{qp},\,{\hat G}_{qp}]    
+[{\hat H}_{pp},\,{\hat G}_{qq}]\right)    
\mea    
\mea    
-\frac{7i\hbar^3}{5760}    
\left(    
[{\hat H}_{qqqq},\,{\hat G}_{pppp}]    
-4[{\hat H}_{qqqp},\,{\hat G}_{qppp}]    
+6[{\hat H}_{qqpp},\,{\hat G}_{qqpp}]    
\right.    
\mea    
\mea    
\left.-4[{\hat H}_{qppp},\,{\hat G}_{qqqp}]    
+[{\hat H}_{pppp},\,{\hat G}_{qqqq}]    
\right)+\dots    
\label{class_evolution}    
\eea    
where the classical Hamiltonian $H(q,\,p)$ is    
represented by the operator ${\hat H}={\cal W}^{-1}(H)$ on Hilbert space.
In (\ref{poisson_evolution}) and (\ref{class_evolution}) we have introduced the notation         
\bea
A_p=\partial A/\partial p\,,\qquad A_q=\partial A/\partial q\,,\qquad A_{qp} =\partial ^2/\partial q \partial p\,,\qquad\qquad
\mea\mea    
{\hat A}_p={\cal W}^{-1}(A_p)=[{\hat q}, {\hat A}]/(i\hbar)\,, 
\quad {\hat A}_q={\cal W}^{-1}(A_q)= [{\hat A}, {\hat p}]    
/(i\hbar)\,,     
\mea\mea    
{\hat A}_{qp} = {\hat A}_{pq}= {\cal W}^{-1}(A_{qp})=
[[{\hat q}, {\hat A}],{\hat p}]/(i\hbar)^2\,, \qquad    
\label{qderivatives}    
\eea    
and so on.    
In (\ref{class_evolution}), the numerical coefficients are those in the expansion of     
$1/(2i\sin(\hbar/2))$ in ascending powers of $\hbar$.   See \cite{Bracken03} for details.      
%Perhaps make a comment about the Bernoulli numbers here?   
  
Thus, to lowest order in $\hbar$, the classical evolution of ${\hat G}$ is given   
by the quantum equation  
  
\bea  
\frac{\partial {\hat G}(t)}{\partial t}=    
\frac{1}{i\hbar}[{\hat H},{\hat G}]\,,  
\label{zeroth_semiquantum}  
\eea  
whereas the first order semiquantum approximation has   
\bea    
\frac{\partial {\hat G}(t)}{\partial t}=    
\frac{1}{i\hbar}[{\hat H},{\hat G}]    
-\frac{i\hbar}{24}\left([{\hat H}_{qq},\,{\hat G}_{pp}]    
-2[{\hat H}_{qp},\,{\hat G}_{qp}]    
+[{\hat H}_{pp},\,{\hat G}_{qq}]\right)\,,  
\label{first_semiquantum}  
\eea  
and so on.

In what follows,    
for simple nonlinear Hamiltonians $H$, we    
consider first and second order semiquantum approximations to the    
classical evolution of ${\hat G}$,     
and look at the resulting effects on expectation values of key observables
by using those approximations to ${\hat G}$ in (\ref{Gproperties}).      
We also consider the behaviour of the    
spectrum of ${\hat G}$ as these successive approximations are introduced.      
    
Finally, we compare these results with    
corresponding results obtained using semiclassical approximations  that    
are determined by considering succesively more terms from   
the well-known series analogous to (\ref{class_evolution}) for    
the evolution of the Wigner function $W(q,\,p,\,t)$, namely \cite{Moyal49}   
\bea    
\frac{\partial W}{\partial t}    
=\{H,W\}_{*}   
=\left(H_q W_p-H_pW_q\right)\qquad\qquad\qquad\qquad    
\mea\mea    
-    
\frac{\hbar^2}{3!2^2}\left(H_{qqq}W_{ppp}-    
3H_{qqp}W_{qpp}+3H_{qpp}W_{qqp}-    
H_{ppp}W_{qqq}\right)\qquad    
\mea\mea    
+\frac{\hbar^4}{5!2^4}\left(H_{qqqqq}W_{ppppp}-5H_{qqqqp}W_{qpppp}+    
10H_{qqqpp}W_{qqppp}-10H_{qqppp}W_{qqqpp}\right.    
\mea\mea    
\left.+5H_{qpppp}W_{qqqqp}-H_{ppppp}W_{qqqqq}\right)-\dots\,,\qquad    
\label{quant_evolution}    
\eea    
where $\{\,,\,\}_{*}$ denotes the Moyal (star) bracket.    
The numerical    
coefficients in this series are those in the expansion of $2\sin(\hbar/2)/\hbar$     
in ascending powers of $\hbar$. 

In this case, the formulas for the quantum averages, analogous to the classical formulas 
(\ref{Gproperties}),
are 

\bea
\langle {\hat A}\rangle (t)=
{\rm Tr}({\hat \rho (t)}{\hat A})=
\langle  A\rangle (t)
=\int W(q\,,p\,,t)\,A(q\,,p)\,dq\,dp\,,
\label{quant_averages}
\eea
where ${\hat \rho}$ is the quantum density operator, and $A={\cal W}({\hat A})$.
Semiclassical approximations to these averages are obtained by
inserting in (\ref{quant_averages}), successive approximations to $W$
found from (\ref{quant_evolution}).

\section{A class of simple nonlinear systems}    
    
The equations of motion for linear systems (quadratic Hamitonians) in    
the quantum and classical regimes are identical,    
whether represented in the Hilbert space or phase space formulations.   
Differences only arise for     
nonlinear systems which, in most cases, can only be studied numerically.     
    
One class of nonlinear dynamical systems for which some analytic results
can be obtained
are systems for which the Hamiltonian is a polynomial in     
the simple harmonic oscillator Hamiltonian    
\begin{equation}    
H_0 \,=\, p^2/2m + m\omega^2q^2/2\,.   
\end{equation}    
We introduce the set of classical Hamiltonians of the form    
\begin{equation}    
H \,=\, E\sum_{k=0}^K b_k \left(H_0/E\right)^k\,, 
\label{hamiltonian_set}  
\end{equation}    
where the $b_k$ are dimensionless constants, $K$ is a    
positive integer and $E$ is a suitable constant with the dimensions of Energy. 
Hamiltonians of this form may seem somewhat artificial, but   
interactions of this general form play a significant role    
in the theory and simulation of Kerr media and laser-trapped Bose-Einstein condensates \cite{BEC1,BEC2}.   
    
The primary advantages of these Hamiltonians $H$ and their quantizations
${\hat H}={\cal W}^{-1}(H)$    
in the present context are that they generate exactly solvable dynamics in both cases,
and that the classical dynamics and the quantum dynamics
display very different behaviours for $K> 1$, for sufficiently large times
\cite{Milburn86}.

\begin{figure}    
\centerline{\psfig{figure=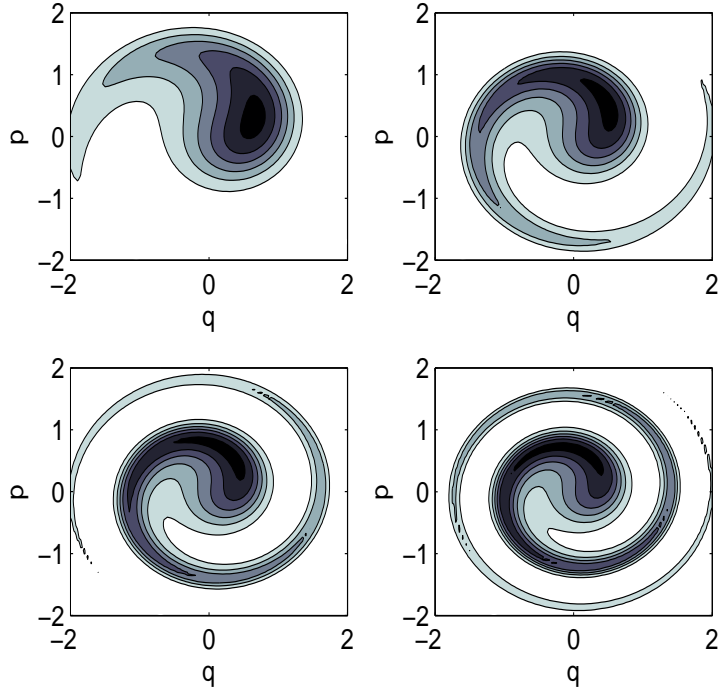,height=100mm,width=110mm}}    
\caption{(Color online) Density plots showing the classical  
evolution of an initial Gaussian density centered at $\alpha_0=q_0=0.5$,  
with $\kappa=2$, as generated by the Hamiltonian $H=H_0^2/E$.  
The parameters $m,\omega,E$ have been set equal to $1$, and the times of the plots are,
from left to right and top to bottom,  
$t=\pi/4$, $t=\pi/2$, $t=3\pi/4$ and $t=\pi$. 
\label{whorl_fig}  
} 
\end{figure}

In order to investigate the classical dynamics generated on phase space by such    
Hamiltonians, we first introduce the    
dimensionless complex conjugate variables 
\bea
\alpha = (\sqrt{m\omega}q+ip/\sqrt{m\omega})/\sqrt{2\hbar}\,,\qquad 
\overline{\alpha} = (\sqrt{m\omega}q-ip/\sqrt{m\omega})/\sqrt{2\hbar}\,,
\label{alpha_alphabar}
\eea  
so that $H_0=\hbar\omega \overline{\alpha}\alpha$. In the classical context,  
$\hbar$ should be considered as simply   
an arbitrary constant with dimensions of action. However, we
think of $E$ as a `classical' energy, and $\hbar\omega$ as a `quantum' energy, 
and define the dimensionless parameter 
\bea
\mu=\hbar\omega/E\,,
\label{Econdition}
\eea  
whose value characterises the relative sizes of 
quantum and classical effects in what follows.
In the `Heisenberg picture' of classical dynamics,
the equation of motion for $\alpha (t)$ is    
\begin{equation}    
\frac{d\alpha}{dt} \,=\, \frac{1}{i\hbar}\frac{\partial H}{\partial    
\overline{\alpha}} \,=\, -i\omega \alpha H'(H_0)\,,    
\end{equation}    
and, since $H_0$ is a constant of the motion,   
\begin{equation}    
\alpha(t)\, =\, \alpha(0)\exp(-i\omega t H'(H_0))\,.    
\end{equation}
If there is a degree of uncertainty in the    
initial conditions of the system, described by a Liouville density    
$\rho(\alpha,\overline{\alpha})$, then the dynamics of the system
may be more conveniently described in 
the `Schr\"odinger picture' of classical dynamics,
where the state at time $t$ is described by the new Liouville 
function $\rho(\alpha,\overline{\alpha},t) = \rho(\alpha(-t),   
\overline{\alpha}(-t))$ \cite{Balescu}.    
When $K > 1$, so that $H$ is a nonlinear function of $H_0$,    
the dynamics leads to rotations in the complex $\alpha,\overline{\alpha}$ plane    
that vary in angular frequency as a function of the radial distance from the origin.    
The effect of this radial variation in the angular frequency 
on an initial density profile localised near some point in the phase plane, is to create   
`whorls' about the origin \cite{Milburn86}  
as time progresses, as shown in Fig. \ref{whorl_fig} for the case $H=H_0^2/E$.    
    
We are interested in the statistical properties of the evolving     
classical state, such as the mean values of position and momentum and    
the variance in these quantities. In general these statistics must    
be determined numerically, but the calculations may be    
simplified for certain classes of Liouville densities.    
We consider Gaussian densities    
\[    
\rho_{\gamma,(q_0,p_0)}(q,p) \,=\, \frac{\omega}{2\pi\gamma^2}   
\exp(-(\tfrac12m\omega^2(q-q_0)^2+\tfrac12(p-p_0)^2/m)/\gamma^2)\,,    
\]    
which can be rewritten as    
\begin{equation}    
\rho_{\kappa,\,\alpha_0}(\alpha) \,=\, \frac{\kappa}{2\pi}    
\exp(-\kappa|\alpha-\alpha_0|^2)\,,    
\label{gaussian_def}   
\end{equation}    
where $\kappa = \hbar\omega/\gamma^2, \quad\alpha_0 = (\sqrt{m\omega}q_0   
+ip_0/\sqrt{m\omega})/\sqrt{2\hbar}$, and the effect of the    
change of measure has been incorporated into the normalisation co-efficient.    
    
When the initial density is of this form,    
the $m$th moment of $\alpha$ at time $t$ is given by    
\bea    
\la \alpha^m\ra (t)\,=\,\frac{\kappa}{\pi} \int_\mathbb{C}    
\alpha^m e^{-i\omega m t H'(\hbar\omega\,\overline{\alpha}\alpha)}  
e^{-\kappa|\alpha-\alpha_0|^2} d^2\alpha\,.    
\label{alpha_moment}
\eea    
This integral can be simplified by setting $\alpha = r\exp(i\phi)/\sqrt{2}$    
and $\alpha_0=r_0\exp(i\phi_0)/\sqrt{2}$ and then evaluating  the integral over    
$\phi$ so that    
\begin{equation}    
\la \alpha^m \ra(t)  \,=\, \frac{\kappa}{2^{m/2}} e^{im\phi_0 -    
\kappa r_0^2/2} \int_0^\infty r^{m+1} \,e^{-\kappa r^2/2}\,   
I_m(\kappa rr_0)e^{-im\omega tH'(\hbar\omega r^2/2)} dr\,, 
\label{reduced_moments}   
\end{equation}    
where $I_m$ denotes the modified Bessel function \cite{Abramowitz} of the first kind of order $m$.    
The classical moments of primary interest are the mean and    
standard deviation in position and momentum.    
These can be constructed from the moments in $\alpha$ and $\overline{\alpha}$ as    
\bea    
\la q \ra(t)  \,=\, \sqrt{(2\hbar/m\omega)}\,    
{\rm Re}\{\la \alpha\ra(t)\}\,, \qquad  \la p \ra (t)  \,=\, \sqrt{2\hbar    
m\omega}\,{\rm Im}\{\la \alpha\ra(t) \} \,,
\mea\mea
\qquad\qquad\Delta q(t) \,=\,\sqrt{\hbar   
(\la |\alpha|^2\ra- {\rm Re}\{\la\alpha\ra(t)^2\})/m\omega  - \la q\ra(t)^2}\,,\qquad\qquad 
\mea\mea       
\qquad\qquad\Delta p(t) \,=\, \sqrt{\hbar m\omega(\la |\alpha|^2\ra -    
{\rm Re}\{\la\alpha\ra(t)^2\}) - \la p\ra(t)^2}\,.\qquad\qquad    
\label{moment_expressions}
\eea    
Note that the moment $\la |\alpha|^2\ra (t) = |\alpha_0|^2 + 1/\kappa$ is a constant of the motion.    
    
If $K>1$, then it follows from (\ref{reduced_moments}) that
$\la\alpha^m\ra(t) \to 0$ as $t\to\infty$ for every $m>0$,   
by the Riemann-Lebesgue Lemma \cite{Churchill}.  In particular, the mean position and momentum tend    
towards zero as time increases. This is in sharp   
contrast to what happens with   
the corresponding quantum evolution, where periodic behaviour
occurs \cite{Milburn86}. 
   
On phase space, the quantum dynamics    
is determined by the Moyal bracket expansion    
(\ref{quant_evolution}),    
from which comparisons can    
be made with the classical dynamics in the phase space setting. 
As already indicated, the transition from classical to quantum  
dynamics can be studied by successively adding on to the classical 
Poisson bracket evolution,  higher order terms in  
that expansion 
\cite{Wigner32}, \cite{Osborn95} until the full quantum dynamics is obtained.  Note that  
for polynomial Hamiltonians such as (\ref{hamiltonian_set}), 
the series (\ref{quant_evolution}) terminates. This process then defines a terminating sequence of 
semiclassical approximations, starting with the classical evolution, and ending with the quantum one.  
  
On the other hand, in order to define semiquantum approximations
to classical dynamics, we represent both the quantum and classical    
dynamics on Hilbert space in terms of the Hamiltonian operator    
$\hat{H}$ which, given (\ref{hamiltonian_set}), takes the form    
\begin{equation}    
\hat{H} \,=\, \MW^{-1}(H)\,=\,E\sum_{k=0}^K c_k \left(\hat{H}_0/E\right)^k\,.
\label{quantum_hamiltonians}    
\end{equation}      
The dimensionless co-efficients  $c_k$ are determined by,    
but are not identical to, the $b_k$ in (\ref{hamiltonian_set}), except that $c_K=b_K$.    
For example if $H=H_0^3/E^2$,    
then  $\hh = \hh_0^3/E^2 + 5(\hbar\omega)^2    
\hh_0/4E^2$
as in(\ref{order_six_hamiltonian}) below.   
The relation between the coefficients $b_k$ and $c_l$ can be determined   
using recurrence relations, but explicit formulas are very complicated. What 
is important to note here is that, with $H$ a `classical' Hamiltonian and
the $b_k$ assumed independent of 
$\hbar$, the $c_k$ for $k>0$ typically have the form $c_k=b_k+{\rm o}(\hbar)$ as $\hbar\to 0$.    
%Should I incorporate these    
    
Since ${\hat H}$ is a function of the oscillator Hamiltonian operator $\hh_0$,  
it can be diagonalised on the well-known  number eigenstates. We introduce the  
creation and annihilation operators $\ha = \MW^{-1}(\alpha)$ and $\had =  
\MW^{-1}(\overline{\alpha})$, and the number operator $\hn = \had\ha$,  
so that $\hh_0 = \hbar\omega(\hn + 1/2)$. Then $\hh$ is diagonal on the  
states $|n\ra$, $n=0,\,1,\,2,\,\dots$ with $\hn\, |n\ra = n|n\ra$, and has eigenvalues given by    

\begin{equation}    
\lambda_n \,=\, E\sum_{k=0}^K c_k \left(\frac{\hbar\omega(n+1/2)}{E}\right)^k\,
=\,E\sum_{k=0}^K c_k \,\mu^k\,(n+1/2)^k\,.
\label{eigenvaluesA}    
\end{equation}

Our object  
is to construct semiquantum approximations by using the expansion  
(\ref{class_evolution}) and, beginning  
with the quantum dynamics,  to successively add on higher order correction  
terms to this until the full classical dynamics is obtained. Because  
the series (\ref{class_evolution}) also terminates for polynomial Hamiltonians, we get in this way a terminating
sequence of semiquantum approximations, beginning with the quantum evolution, and ending
with the classical one. 
 
%In particular, it  
%is of interest to gauge how well expressions involving a single correction  
%term to the quantum evolution, approximate the full classical evolution.    

\section{Example 1: Hamiltonian of degree 4}

The first example that we consider has 
\bea
H = H_0^2/E,\qquad  \hat{H} = \hat{H}_0^2/E +  
(\hbar\omega)^2/4 = \mu\hbar\omega (\hn^2+\hn +1/2)\,.
\label{quartic_hamiltonian}
\eea
Since $H$ is quadratic in $H_0$, and hence quartic in $q$ and $p$,
the dynamics is  nonlinear.   
It is then to be expected that the classical and quantum evolutions produce comparable values
and behaviours of
the expectation values of observables 
for only a limited period  
of time, sometimes referred to  
as the break time \cite{Breaktime}. This depends on $\hbar$ and the properties of the initial state. 
The numerical value attributed to the 
break time in a particular case  
is dependent on which criterion is used to determine differences between  
the quantum and classical dynamics, so it should be interpreted only as a guide  
to the time-scale over the which the evolutions produce similar 
values for observable quantities.  

\begin{figure}    
\centerline{\psfig{figure=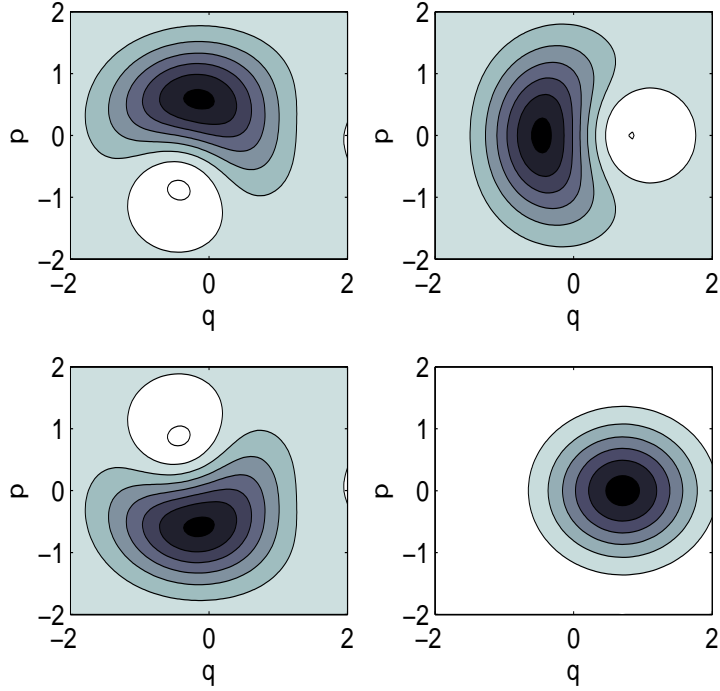,height=100mm,width=110mm}}    
\caption{(Color online) Quantum evolution of  
an initial Gaussian Wigner function, with the same parameter values used  
in Figure \ref{whorl_fig}, and shown at the same times. Regions on which the
pseudo-density becomes negative are shown in white.
\label{quant_gaussian_fig}}  
\end{figure}

Differences between the classical and quantum evolutions are already apparent
if we compare the classical picture in Fig.\ref{whorl_fig}, with 
the quantum evolution of the Wigner function on phase space in Fig.\ref{quant_gaussian_fig}, which starts
with the same initial Gaussian density, and covers the same length of time.  In the first place, the Wigner function immediately
develops negative values on some regions, shown in white in Fig.\ref{quant_gaussian_fig}.  Secondly, the
quantum evolution is 
periodic, unlike the classical evolution, with period $2\pi/\mu\omega$ in this case.  This can be seen more clearly from the
formula (\ref{quartic_elementsA}) below for the matrix elements of the density operator, all of which have this period. Many studies
in recent years have explored in detail such characteristic differences between quantum and classical dynamics, 
especially in the context of 'chaos'\,; see for example \cite{Berry77,Gutzwiller90}.

In this first example, the full classical evolution on Hilbert space is given by adding just  
the first correction to the quantum evolution in (\ref{class_evolution}), and hence in this case  
there are only the zeroth order semiquantum dynamics and the full classical dynamics
to study. Similarly, the full quantum evolution 
on phase space is given with just the first correction to the classical evolution in (\ref{quant_evolution}), 
so there are only zeroth order semiclassical dynamics and the full quantum dynamics to study.  
In short, there is no dynamics 
`in between' the classical and quantum dynamics in this case.  Nevertheless, the simplicity  
of this example makes it relatively easy to derive analytic expressions  
for the classical dynamics on Hilbert space that are relevant to more complicated cases. 
    
The quantum evolution in Hilbert space $\MH$ in this example is determined by    
\begin{equation}    
\frac{d \hat{\rho}}{dt} \,=\, \frac{(\hbar\omega)^2}{i\hbar E} [\hn(\hn+1),\hat{\rho}]
=-i\mu\omega[\hn(\hn+1),\hat{\rho}]\,.
\label{quartic_quant_evolutionA}    
\end{equation}    
If we represent $\hat{\rho}$ in the number basis, its matrix elements 
$\rho_{nm} \,=\, \la n | \hat{\rho} | m \ra$  
then evolve as    
\begin{equation}    
\rho_{nm}(t) \,=\, e^{-it(E_n-E_m)/\hbar} \rho_{nm}(0)\,,
\label{quartic_elementsA}    
\end{equation}    
where $E_n$, $E_m$ are eigenvalues of $\hat{H}$ with, from (\ref{eigenvaluesA}),      
$E_n = \mu\hbar\omega (n^2+n+1/2)$. 
    
Changing variables from $(q,p)$ to $(\alpha, \overline{\alpha})$, we find 
the classical evolution (\ref{class_evolution}),
on $\MH$ in this example as     
\bea    
\frac{d \hat{G}}{dt} \,=\, -i\mu\,\omega\, [\hn(\hn+1),\hat{G}]\qquad\qquad  
\mea\mea
-\frac{1}{24 i\hbar} \left( [\hhaa,\hgbb] -2[\hhab,\hgab] + [\hhbb,\hgaa] \right)\,, 
\label{quartic_class_evolutionA}   
\eea    
where $\hh_{\alpha\alpha} = \MW^{-1}(\partial^2 H/\partial \alpha^2),\,  
\hh_{\alpha\overline{\alpha}} = \MW^{-1}(\partial^2 H/\partial  
\alpha\partial\overline{\alpha})$, {\em etc.} In this expression the correction term  
appears to be of lower order in $\hbar$ than the zeroth order term because we  
have introduced the operator $\hn$ which is ${\rm O}(1/\hbar)$, and 
derivatives with respect to $\alpha$ and $\overline{\alpha}$  
which are ${\rm O}(\sqrt{\hbar})$. 
After some algebraic manipulation, (\ref{quartic_class_evolutionA}) simplifies to
\begin{equation}    
\frac{d \hat{G}}{dt} \,=\, -i(\mu\omega/2) 
\left( [\hn(\hn+1), \hg] + [\hn, \ha\hg\had + \had\hg\ha] \right)\,. 
\label{quartic_class_evolutionB}   
\end{equation}
Note that although as remarked above this must be fully equivalent to the classical evolution
of the Liouville density, there is still dependence on $\hbar$ in the RHS, because of the dependence
implicit in the definition of ${\hat G}$.     
The effect of the evolution  (\ref{quartic_class_evolutionB}) can be made clearer if we write $\hat{G}$  
as the `operator vector' $|G\ra\ra \in \MH\otimes\MH^*$ \cite{Muga92} and  
replace the right and left action of operators on $\hg$ by superoperators \cite{Muga93,Muga94,Vercin00} acting on $|G\ra\ra$, labelled
by the subscripts $r$  
and $l$ respectively, {\it i.e.} $\hat{A}_l|\hg\ra\ra \equiv \hat{A}\hat{G}$  
and $\hat{A}_r|\hg\ra\ra \equiv \hat{G}\hat{A}^\dagger$. We can then  
rewrite (\ref{quartic_class_evolutionB}) in the operator vector form    
\begin{equation}    
\frac{d |G\ra\ra}{dt} \,=\, -i(\mu\omega/2)  
(\hn_l-\hn_r)\left(\hn_l+\hn_r+1+\ha_l\ha_r + \had_l\had_r\right)|G\ra\ra \,. 
\label{quartic_class_evolutionC}   
\end{equation}    
If we now define the new superoperators 
\bea
\hn_-=\hn_l-\hn_r\,, \qquad\hat{X}_1 =  
(\hn_l+\hn_r+1)/2\,, 
\mea\mea
\hat{X}_2 = (\ha_l\ha_r + \had_l\had_r)/2\,,\qquad  
\hat{X}_3=(\ha_l\ha_r - \had_l\had_r)/2\,, 
\label{sl2operators}
\eea
then the $\hat{X}_i$ close to form an     
$sl(2)$ algebra of superoperators that commute with $\hn_-$ and satisfy on $\MH\otimes \MH^*$ 
the commutation relations    
\begin{equation}    
[\hat{X}_2,\hat{X}_1]=\hat{X}_3\,, \quad [\hat{X}_3,\hat{X}_2]= 
\hat{X}_1\,, \quad [\hat{X}_3,\hat{X}_1]=\hat{X}_2\,.
\label{sl2relations}    
\end{equation}    
In terms of these new superoperators, the classical evolution of  
$|G\ra\ra$ is given by    
\begin{equation}    
d |G\ra\ra \,/dt \,=\, -i\mu\,\omega\,\hn_-(\hat{X}_1+\hat{X}_2)|G\ra\ra\,.
\label{quartic_sl2evolution}    
\end{equation}    
    
Since the $sl(2)$ superoperators commute with $\hn_-\,$, one method of  
simplifying the evolution is to decompose the tensor product space  
$\MH\otimes \MH^*$ into representations of $sl(2)$ labelled  
by the eigenvalues of $\hn_-\,$. These eigenvalues may take any integer  
value $\nu \in \mathbb{Z}$ and the corresponding $sl(2)$ lowest-weight  
operator eigenvector of ${\hat X}_1$
is given by $|\nu,0\ra\ra$ for $\nu\geq 0$ and $|0,-\nu\ra\ra$ for  
$\nu <0$. Here we have introduced the basis of operator vectors  
$\{|n,m\ra\ra,\, n,m= 0,1,2,\ldots \}$ in $\MH\otimes\MH^*$, that corresponds to the  
basis of operators $\{|n\ra\la m|,\,n,m= 0,1,2,\ldots\}$ in the space of operators acting on  $\MH$,
with $\hn_l |n,m\ra\ra =  
n|n,m\ra\ra$ and $\hn_r |n,m\ra\ra = m|n,m\ra\ra$.     
    
Now we can write

\bea
|G\ra\ra =
\sum_{\nu\in \mathbb{Z}}\,|G^{(\nu)}\ra\ra\,,    
%\bigoplus_{\nu\in \mathbb{Z}} |G^\nu\ra\ra\,, 
\label{sl2expand}
\eea
where 
\begin{equation}
|G^{(\nu)}\ra\ra=\left\{
\begin{array}
{r@{\quad }l}
\sum_{n=0}^{\infty} G_n^{(\nu)} |n+\nu,n\ra\ra & ,\quad\nu\geq 0\\ \\
\sum_{n=0}^{\infty} G_n^{(\nu)} |n,n-\nu\ra\ra  & ,\quad\nu<0
\end{array}
\right.
\end{equation}
with
\bea
G_n^{(\nu)} = \la\la n+\nu,  
n|G\ra\ra =\la n+\nu|\hg|n\ra \quad {\rm for}\quad \nu\geq 0\,,
\mea \mea
G_n^{(\nu)} = \la\la n,n -\nu|G\ra\ra =\la n|\hg |n-\nu\ra \quad {\rm for}\quad \nu<0\,.
\label{Gcoeffs}
\eea    
    
The classical evolution (\ref{quartic_sl2evolution}) of $|G\ra\ra$ generated by $\hh$ 
now leads to    
\begin{equation}    
|G(t)\ra\ra \,=\, \sum_{\nu \in \mathbb{Z}} |G^\nu(t)\ra\ra\,=\,   
\sum_{\nu \in \mathbb{Z}} \exp(-i\nu\mu\omega\, t  \hat{P}^{(\nu)}) |G^{(\nu)}(0)\ra\ra\,,
\label{quartic_class_evolutionD}    
\end{equation}    
where  the only nonzero matrix elements of the superoperator $\hat{P}^{(\nu)}$ are given by    
\bea    
(\hat{P}^{(\nu)})_{nn}\, =\, n+(|\nu|+1)/2\,,\,\qquad\qquad\qquad\qquad 
\mea\mea
(\hat{P}^{(\nu)})_{nn+1} 
\, =\, (\hat{P}^{(\nu)})_{n+1n} \,=\, \sqrt{(n+1)(n+|\nu|+1)}\,/2\,. 
\label{quartic_matrix_elements}   
\eea    
As a counterpoint to the quantum evolution (\ref{quartic_elementsA}) 
of the matrix elements of $\hr$,  
we find that the classical evolution transforms the matrix elements of $\hg$
along each diagonal independently:    
\begin{equation}    
G_{nm}(t) \,=\, \sum_{r=0}^\infty
\left(\exp[-i(n-m)\mu\omega t\hat{P}^{(n-m)}]\right)_{mr}\,G_{n-m+r\,r}(0) \,,
\label{quartic_elementsB}    
\end{equation}    
for $n\geq m$. By the $\nu$th diagonal, we mean the set of matrix elements  
for which  the row label minus the column label is equal to  
$\nu$. If $\hg$ is initially Hermitian, then it remains Hermitian under  
this evolution and the matrix elements of $\hg(t)$ above the main diagonal  
can be obtained by complex conjugation from those below.    
    
We can deduce at once several properties of the classically evolved matrix $\hg(t)$  
by recalling some properties of  the classical evolution on the phase plane. Firstly, the total  
amount of probability on the plane remains constant, which implies that  
the trace of $\hg(t)$ should also remain constant. That this is reproduced in  
the Hilbert space analysis can be seen from (\ref{quartic_elementsB}) by setting $n=m$ and  
observing that the diagonal elements of $\hg$ are constants of the motion.  
Secondly, the integral over phase space of the square of the Liouville density  
is also a constant of the motion, and this is equivalent to the Hilbert space condition  
that $\Tr(\hg(t)^2)$ should remain constant. This property of $\hg(t)$  
can be verified from (\ref{quartic_elementsB}) by noting that the evolution of each diagonal of $\hg (t)$  
is unitary and hence that the sum of the squares of the elements of the $\nu$th diagonal  
is a constant of the motion. It then follows that the sum over all diagonals,  
equal to $\Tr(\hg(t)^2)$, is also a constant.    
    
Note however that the classical evolution generated by $\hh$ is not unitary in the sense  
one uses when describing quantum dynamics, that is to say, in the Hilbert space $\MH$. 
Instead, it defines an  
evolution of $\hg$ that corresponds to a unitary superoperator evolution of the operator vector  
$|\rho\ra\ra$ in $\MH\otimes\MH^*$. This has the consequence that some features of a quantum evolution remain,  
such as the trace relations described above, but that other  
characteristic features of a quantum evolution are lost. For example,  
there is no requirement for the eigenvalues of the Groenewold operator  
to remain constant under the motion. These eigenvalues can and do change and  
in general, negative eigenvalues will typically develop over time, even if all
eigenvalues of $\hg$ are intially non-negative.  
    
When examining the relationship between quantum and classical dynamics, it  
is primarily of interest to consider the limit $\hbar \rightarrow 0$ in some appropriate way.  
There is some flexibility in how to do this. In \cite{Milburn86}, 
the limit was taken in such a way that the  
initial phase space density approaches a delta-function in $q$ and $p\,$,  
so that both the quantum and classical dynamics approach that of a classical  
trajectory. In this paper, we shall instead investigate numerically 
the effect of reducing the value  
of $\hbar$ in such a way that the initial Gaussian phase 
space density is kept unchanged.    
    
This involves a scaling of three key parameters: $\kappa$, $\mu$ and $\alpha_0$.  
The first two of these depend linearly on $\hbar$, while 
$\alpha_0$ is linear in $(1/\sqrt{\hbar})$.  
In order to keep the initial density constant as $\hbar \rightarrow 0$,  
we must have $\kappa$, $\mu \rightarrow 0$ and $\alpha_0 \rightarrow \infty$,  
while keeping $\kappa/\mu$ and $\kappa|\alpha_0|^2$ constant.     
    
\begin{figure}    
\centerline{\psfig{figure=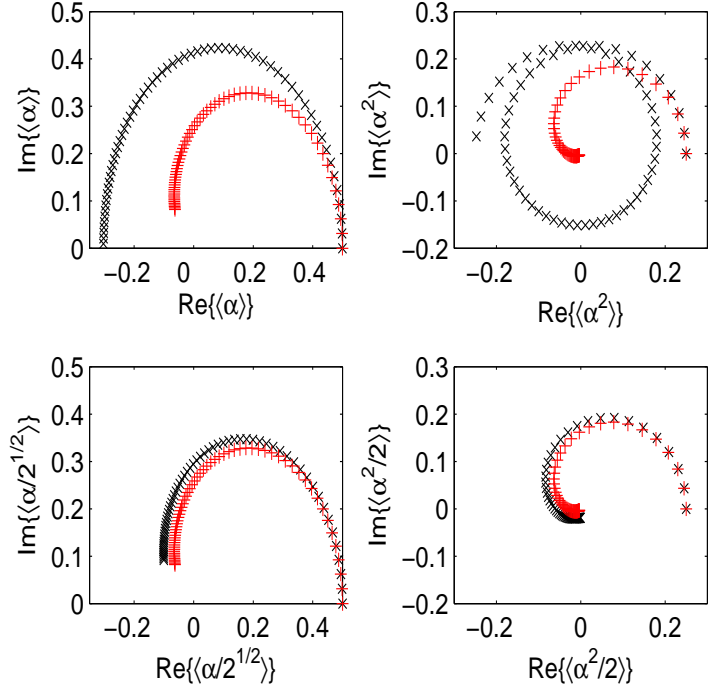,height=100mm,width=110mm}}    
\caption{(Color online) Classical and quantum evolution of the first  
and second moments of $\alpha=(q+ip)/\sqrt{2}$ as generated by $H=H_0^2/E$.  
Points on the classical curves are marked by + and points on the quantum  
curves by x. Again $m=\omega=E=1$,  
and here $q_0=\sqrt{2},p_0=0$. The first moments are graphed on the left and the second moments on the right.  
In the two upper graphs, $\kappa=2,\mu=1/2, 
\alpha_0=1/2$ and in the lower two graphs, $\kappa=1,\mu=1/4, \alpha_0=1/\sqrt(2)$.
\label{quartic_moments}}  
\end{figure}

In Fig.\ref{quartic_moments}, we graph the first and second moments  
of $\alpha$ for two sets of the parameters $\kappa$, $\mu$ and  
$\alpha_0$, with $t$ in the range $[0,\pi]$.  For the top two graphs in Fig.\ref{quartic_moments},  
the values $\kappa=2$, $\mu=1/2$ and $\alpha_0=1/2$ are used, while in  
the bottom two graphs, the values $\kappa=1,\mu=1/4$ and $\alpha_0=1/\sqrt(2)$ are used,
corresponding to $\hbar\rightarrow\hbar/2$.  
The first moments are graphed on the left in Fig.\ref{quartic_moments}, and it can be seen that the  
quantum evolution lies  close to the classical for much longer with the smaller (effective) value of $\hbar$.  
A more exaggerated sign of this convergence can be seen in the graphs  
of the second moments on the right. In the upper graph, one can see that the  
classical moment approaches the origin, whereas the quantum moment  
exhibits a symmetry in $q$. This symmetry is a sign of the recurrence of  
the initial state at $t=2\pi$. In the lower graph however, the recurrence is not  
due to occur until $t=4\pi$ and the quantum evolution 
resembles the classical much more closely for $t\in [0,2\pi]$.    
    
\section{Example 2: Hamiltonian of degree 6}    
In order to compare and contrast semiquantum and semiclassical  
dynamics, we need to select a Hamiltonian of least degree 3  
in $H_0$, so that there are nontrivial semiclassical and semiquantum approximations 
lying between classical and quantum dynamics. The simplest such case has 
\bea
H=H_0^3/E^2\,,\quad\hh &=& \hh_0^3/E^2+5(\hbar\omega)^2\hh_0/4E^2
\mea\mea
&=&\mu^2\,\hbar\,\omega\,(\hn^3+3\hn^2/2+2\hn+3/4)\,.
\label{order_six_hamiltonian}
\eea    
    
The eigenstates of $\hat{H}$ are again the number states  
and it is natural to represent the dynamical evolution in Hilbert space 
in this basis. The quantum evolution of the matrix  
elements of an initial density operator $\hr(0)$ on $\MH$ 
is again given by (\ref{quartic_elementsA}), where in  
this case the energy eigenvalues take the form $E_n = \mu^2\hbar\omega(n^3+3n^2/2+2n+3/4)$. 
From this we see that the quantum evolution is again periodic, but now with period
$4\pi/\mu^2\omega$.   
    
The equations governing the classical evolution of the Groenewold operator
on $\MH$ are obtained as before by inserting  
the Hamiltonian into equation (\ref{class_evolution}). The result is that we obtain an  
expression for the full classical evolution that 
starts with the quantum evolution and adds two correction terms:    
\begin{eqnarray}    
\frac{d\hg}{dt}= -i\mu^2\hbar\omega \left[ \hn^3+3\hn^2/2+2\hn, 
\hg\right]\qquad\qquad\quad
\mea\mea
 -\frac{1}{24 i\hbar} \left( [\hhaa,\hgbb] -2[\hhab,\hgab] + [\hhbb,\hgaa] \right)\qquad\qquad    
\mea\mea 
+ \frac{7}{5760 i\hbar} \left( -4[\hhaaab, \hrabbb] +  
6[\hhaabb,\hgaabb] -4[\hhabbb,\hgaaab] \right)\,.
\label{semiquantumB}    
\end{eqnarray}     
Note that terms involving fourth derivatives with respect to  
$\alpha$ and $\overline{\alpha}$ do not appear in   
the second classical correction, because $\partial^4 H/\partial \alpha^4$  
and $\partial^4 H/\partial \overline{\alpha}^4$ vanish in this example.    
    
The correction terms can be simplified by straightforward  but lengthy
algebraic manipulation. Of particular interest is the semiquantum  
evolution equation that includes the quantum evolution and the  
first classical correction. This equation simplifies to    
\bea    
\frac{d\hg}{dt}= -i(\mu^2\,\hbar\,\omega/4)\left(5[\hn^2,\had\hg\ha] + 5[\hn(\hn+2), 
\ha\hg\had]\right.
\mea\mea
\left. - [\hn(\hn+1)(\hn+2),\hr] - [\hn,\ha^2\hr\ha^{\dagger 2} +  
\ha^{\dagger 2}\hr\ha^2 +3\hn\hr\hn]\right)\,.
\label{semiquantumC}    
\eea

If we make use of the $sl(2)$ operators (\ref{sl2operators}) introduced earlier,  
then the corresponding evolution equation for the operator vector $|G\ra\ra$ is given 
in this case by    
\begin{equation}    
d|G\ra\ra/dt \,=\, -i\mu^2\omega\hn_- 
\left( 3(\hat{X}_1\hat{X}_2+\hat{X}_2\hat{X}_1)/2-(\hat{X}_1-\hat{X}_2)^2\right)|G\ra\ra\,.
\label{sl2evolutionB}    
\end{equation}    
The solution of this equation has similar properties to the solution of (\ref{quartic_sl2evolution}). 
In particular, it can again be  
decomposed as in (\ref{sl2expand}), leading in this case to an expression of the form
\begin{equation}    
|G(t)\ra\ra \,=\, \sum_{\nu \in \mathbb{Z}} |G^\nu(t)\ra\ra\,=\,   
\sum_{\nu \in \mathbb{Z}} \exp(-i\nu\mu^2\omega t \hat{Q}^{(\nu)}) |G^\nu(0)\ra\ra
\label{sextic_Gexpand}    
\end{equation}
in place of (\ref{quartic_class_evolutionD}).    
However in this case, 
%the matrix $\hat{Q}^{(\nu)}$, replacing the matrix $\hat{P}$ of the previous case,
%has non-zero matrix elements given by    
%Only got half what I need here - presumably means that I didn't include some terms:)    
%\begin{eqnarray*}    
%\hat{Q}^{(\nu)}_{ nn} &=& 3(n^2+n(|\nu|+1))/2+(|\nu|+1)(|\nu|+3)/4\,,\\    
%\hat{Q}^{(\nu)} _{n+1\,n} \,=\, \hat{Q}^{(\nu)}_{ n\,n+1} &=& (n+1+|\nu|/2)\sqrt{(n+1)(n+|\nu|+1)}/2\,, \\    
%\hat{Q}^{(\nu)}_{ n+2\,n} \,=\, \hat{Q}^{(\nu)}_ {n\,n+2} &=& \sqrt{(n+1)(n+2)(n+|\nu|+1)(n+|\nu|+2)}/4\,.    
%\end{eqnarray*}    
truncation methods for determining the eigenvalues of $\hat{Q}^{\nu}$ appear  
to fail, and hence for computational purposes, it is useful first to introduce  
the unitary $sl(2)$ operator $\hat{U} = \exp(\log(7/3)\hat{X}_3/4)$ 
which enables us to simplify the expression in (\ref{sl2evolutionB}), as     
\bea    
\hat{U}\,\left(3(\hat{X}_1\hat{X}_2+\hat{X}_2\hat{X}_1)/2-(\hat{X}_1-\hat{X}_2)^2\right) 
\,\hat{U}^\dagger = \sqrt{21}\,\left(\hat{X}_1\hat{X}_2+\hat{X}_2\hat{X}_1\right)\,. 
\label{simplification}   
\eea 
This is seen after noting from (\ref{sl2operators}) and (\ref{sl2relations}) that
\bea
3(\hat{X}_1\hat{X}_2+\hat{X}_2\hat{X}_1)/2-(\hat{X}_1-\hat{X}_2)^2&=&
3{\hat X}_+^2/4-7{\hat X}_-^2/4\,,
\mea\mea
{\hat X}_1{\hat X}_2
+
{\hat X}_2{\hat X}_1 &=& ({\hat X}_+^2-{\hat X}_-^2)/4\,,
\label{Xrelations}
\eea
where

\bea
{\hat X}_{\pm}={\hat X}_2\pm{\hat X}_1\,,
\qquad
{\hat X}_3\,{\hat X}_{\pm}={\hat X}_{\pm}({\hat X}_3\pm 1)\,.
\label{rewrite_lie_algebra}
\eea    
The transformation (\ref{simplification}) maps $\hat{Q}^{(\nu)}$ to a tri-diagonal matrix $\hat{Q}^{(\nu)'}$,  
whose only non-zero matrix elements are given by    
\begin{equation}    
\hat{Q}^{(\nu)'}_{n+1\,n} \,=\, \hat{Q}^{(\nu)'}_{ n\,n+1} \,=\, 
\sqrt{21}(n+1+|\nu|/2)\sqrt{(n+1)(n+|\nu|+1)/2}\,,    
\end{equation}    
and for which the eigenvalues can easily be determined numerically.  
The expression for the individual matrix elements of the Groenewold operator in this
semiquantum approximation is now given by    
\begin{equation}    
G_{nm}(t) \,=\, \sum_{r,s,t=0}^\infty U^{(n-m)}_{mr}\exp\left(-i(n-m)\mu^2\omega t 
\hat{Q}^{(n-m)'}\right)_{rs}U^{(n-m)\dagger}_{st}G_{n-m+t,t}(0)\,, 
\label{sextic_classical_elementsA}   
\end{equation}    
where $U^{(\nu)}_{nm} = \la n+\nu,n| \hat{U}|m+\nu,m\ra$ for $\nu\geq 0$.    
    
In order to determine the full classical evolution on Hilbert space,  
we need to evaluate both correction terms in (\ref{semiquantumB}).  
This leads in place of (\ref{quartic_sl2evolution}) to a comparatively simple expression 
for the evolution of the operator vector $|G\ra\ra$:    
\begin{equation}    
\frac{d|G\ra\ra}{dt} \,=\, -3i\omega\mu^2/4 
\hn_-(\hat{X}_1+\hat{X}_2)^2|G\ra\ra\,.
\label{classical_groenewold_evolution}    
\end{equation}
    
Note that this evolution again depends on the operator sum $\hat{X}_1+\hat{X}_2$.  
This property turns out to be shared by the classical evolution for  
each Hamiltonian of the form (\ref{hamiltonian_set}).  The formula corresponding 
to (\ref{classical_groenewold_evolution})
in the general case is    
\begin{equation}    
\frac{d|G\ra\ra}{dt} \,=\, -i\omega\hn_-\sum_{k=0}^N \frac{(k+1)c_k\mu^k}{2^k}  
(\hat{X}_1+\hat{X}_2)^k |G\ra\ra\,. 
\label{general_groenewold_evolution}    
\end{equation}    
However, expressions for the successive semiquantum approximations to the  
quantum evolution, such as (\ref{sl2evolutionB}), are not so easy to
obtain for a general Hamiltonian of the 
form (\ref{hamiltonian_set}).

Returning to the example at hand, with ${\hat H}$ as in (\ref{order_six_hamiltonian}),
the classical evolution of the matrix elements of $\hg$ is given by    
\bea
G_{nm}(t) \,=\, \sum_{r=0}^\infty \left(\exp[-i(n-m)\mu^2\omega t 
\left({\hat P}^{(n-m)}\right)^2]\right)_{mr}G_{n-m+r,r}(0)\,,
\label{sextic_classical_elementsB}    
\eea
with ${\hat P}^{(\nu)}$ as in (\ref{quartic_matrix_elements}).  Eqn. (\ref{sextic_classical_elementsB}) is to be compared 
with (\ref{quartic_elementsB}) in the previous section.    
Working from (\ref{sextic_classical_elementsA}) and (\ref{sextic_classical_elementsB}), 
we can evaluate and compare various quantities of interest
in the semiquantum   
and classical evolutions, such as the expectaton values
of $\alpha$, and the eigenvalues of the Groenewold operator. 

There is also a semiclassical approximation   
to be studied in this case. The classical dynamics involves two 
correction terms to the quantum   
evolution, and the quantum dynamics involves
two correction terms to the classical evolution. Just as the semiquantum dynamics
includes the first correction to the quantum dynamics but not the second, so
the semiclassical dynamics   
includes the first correction to the classical dynamics but not the second.
It is of particular interest to determine   
differences between the semiquantum and the semiclassical dynamics.  

\begin{figure}    
\centerline{\psfig{figure=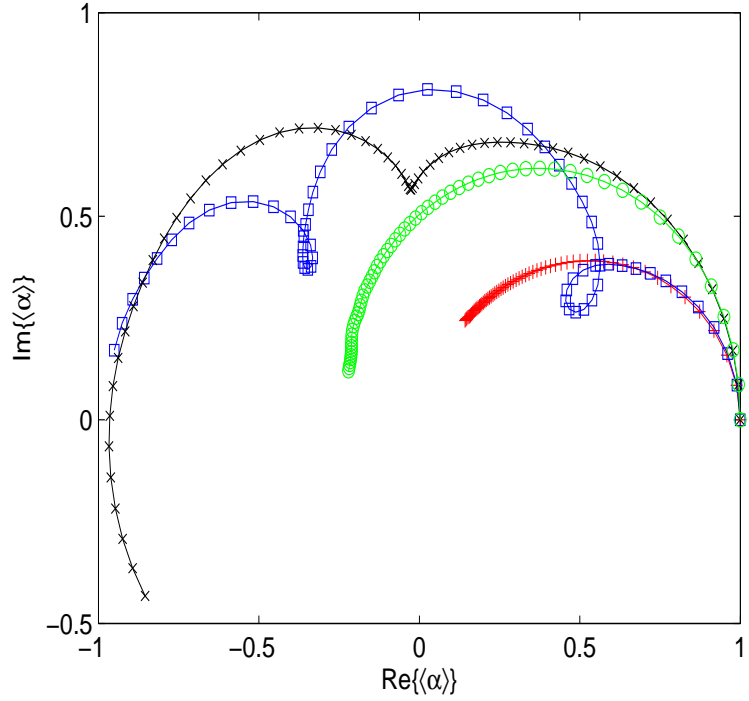,height=100mm,width=110mm}}    
\caption{(Color online) Comparison of first moments of $\alpha=(q+ip)/\sqrt{2}$ for classical,  
semiquantum, quantum and semiclassical evolutions generated by the  
Hamiltonian $H=H_0^3/E^2$. Points on the classical, quantum, semiclassical  and semiquantum curves are  
labelled by +, x, o and $\square$, respectively.  
The evolution is  
over the time-interval $[0,\pi]$ and again $m=\omega=E=1$, with $\kappa=2,\mu=1/2,\alpha_0=0.5$.
\label{sextic_alphaA}}  
\end{figure}

In Fig.(\ref{sextic_alphaA}), we show phase space 
plots of the evolution of the expectation value of $\alpha$ in the semiclassical and semiquantum approximations,
along with the classical  
and quantum evolutions.  
The parameter values were chosen to be $\kappa=2$, $\mu=1/2$ and $\alpha_0=1/2$, and  
the time evolution is over the interval $[0,\pi]$. It is immediately clear  
that the semiclassical and semiquantum approximations are quite distinct.  
For small values of $t$, they closely approximate the quantum and classical 
moments respectively. 
However, the long-term behaviour is very different, as the semiquantum moment  
begins to oscillate in a fashion that is qualitatively similar to the quantum moment,  
whereas the semiclassical moment appears to approach smoothly a fixed constant,  
just as is the case for the classical moment.

\begin{figure}    
\centerline{\psfig{figure=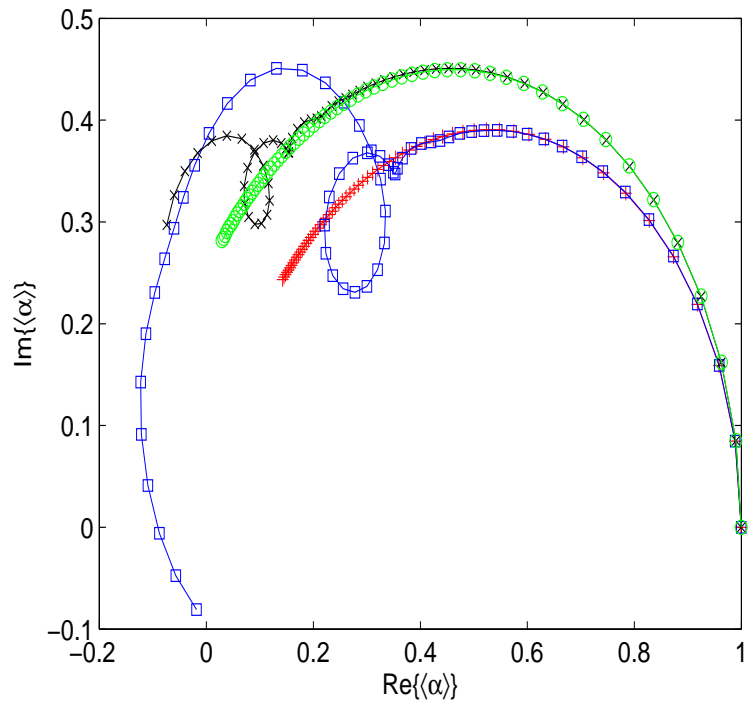,height=100mm,width=110mm}}    
\caption{(Color online) Identical to the preceding Figure except that $\hbar$ has been  
reduced by a factor of two, by setting $\kappa=1,\mu=1/4,\alpha_0=1/\sqrt{2}$.
\label{sextic_alphaB}}  
\end{figure}

How does the behaviour change as $\hbar$ decreases? To examine this,  
we graph in Fig.(\ref{sextic_alphaB}) the first moment in $\alpha$ over $t\in [0,\pi]$,  
but this time with the parameter choices $\kappa=1,\mu=1/4,\alpha_0=1/\sqrt{2}$,  
which is equivalent to dividing $\hbar$ by two. The first thing one observes is  
that the quantum evolution is now much closer to the classical, although it begins  
to oscillate near $t=\pi$. It is also apparent that the semiquantum and  
semiclassical curves are good approximations to the classical and quantum  
moments respectively over most of the time-interval. One can also add that the  
semiclassical moment appears to be a good approximation to the quantum moment  
for a longer time than the semiquantum moment is a good  
approximation to the classical moment.

All the forms of dynamics can 
be studied in phase space or in Hilbert space.  However, 
it must be emphasized that here as in general, 
the zeroth order semiclassical approximation to
quantum dynamics, namely classical dynamics itself, 
when applied in Hilbert space with an initial Groenewold operator ${\hat G}(0)$ 
that is the image ${\cal W}^{-1}(2\pi\hbar\rho)$ of an initial  Liouvillean density,
will in general produce a ${\hat G}(t)$ that is not positive-definite for times $t>0$, even
if ${\hat G}(0)$ is positive-definite, which it need not be. 
Closely related observations have been made in earlier studies of the
evolution of density operators under classical dynamics \cite{Muga92,Muga93,Muga94,Habib02}.  
 
Conversely, the zeroth order semiquantum dynamics,
namely quantum dynamics itself, when applied with an initial Wigner function $W(q,\,p,\,0)$
that 
is the image ${\cal W}({\hat \rho}/2\pi\hbar)$ of an initial quantum density operator,
will in general produce a $W(q\,,p\,,t)$ that is not positive-definite for times $t>0$, even if
$W(q,\,p,\,0)$ is positive-definite, which it need not be.  
    
\begin{figure}    
\centerline{\psfig{figure=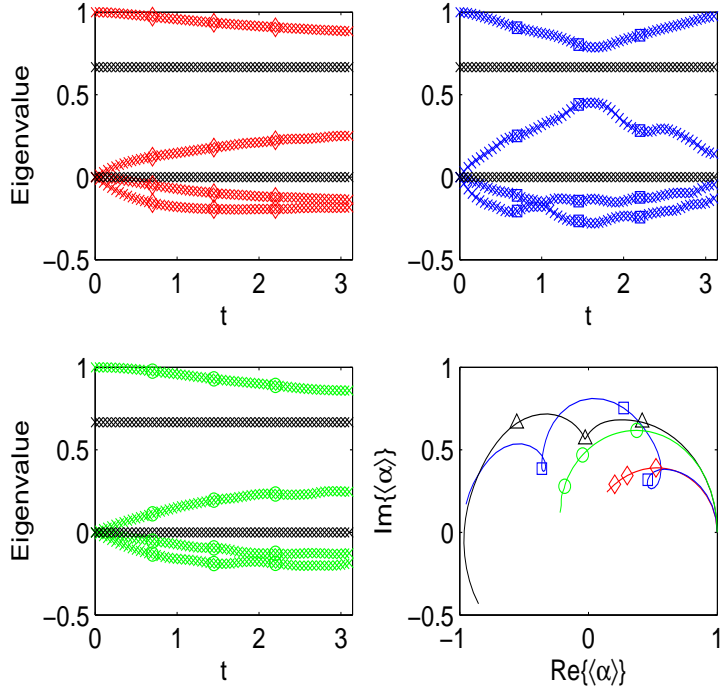,height=100mm,width=110mm}}    
\caption{(Color online) Comparison of largest two and least two eigenvalues for, from
left to right and top to bottom, classical,  
semiquantum  and semiclassical evolutions generated by $H=H_0^3/E^2$,  
for the time interval $[0,\pi]$, and with $m=\omega=E=1$ and $\kappa=2,\mu=1/2,\alpha_0=0.5$.  
The evolution of the first moment is reproduced from Fig.\ref{sextic_alphaB} in the graph at bottom right for comparison.  
Each of the other graphs also features the quantum spectrum $\{0,1\}$ and in all graphs  
the values at the time-points $t=1,2,3$ are marked $\lozenge$ (classical), 
o (semiclassical), $\triangle$ (quantum) and $\square$ (semiquantum).  
\label{sextic_eigenvalues}}  
\end{figure}

The preceding figures use measures of comparison that are natural in the phase space  
formulation of all the dynamics. Alternatively, one can use measures  
that are typically associated with formulations on Hilbert space. In particular,  
one can examine the evolution of the eigenvalues of the density operator
(or the Groenewold operator) in  
the classical, quantum, semiclassical and semiquantum approximations. 
In Fig.(\ref{sextic_eigenvalues}) we show the evolution of the largest pair and smallest pair of eigenvalues 
for the  classical,   
semiquantum and  semiclassical cases. In the quantum case,
the eigenvalues are $0$ and $1$ at all times as the system is in a pure (coherent) state initially, 
and stays in a pure state at all subsequent times. 
The parameter choices and time  
interval used in Fig.(\ref{sextic_eigenvalues}) correspond to those used in Fig.(\ref{sextic_alphaA}). 
We also single  
out the points at which $t=1,2,3$ on each graph for the purpose of comparison with the  
moment curve of Fig.(\ref{sextic_alphaA}) which is reproduced in Fig.(\ref{sextic_eigenvalues}). 
From these graphs, we observe that the  
semiquantum approximation to the classical moment is reasonable up till about $t=1$,  
by which time the eigenvalues of the classical and semiquantum Groenewold operators  
differ to a significant degree. The semiquantum eigenvalues  
appear to display periodic behaviour.   

\begin{figure}    
\centerline{\psfig{figure=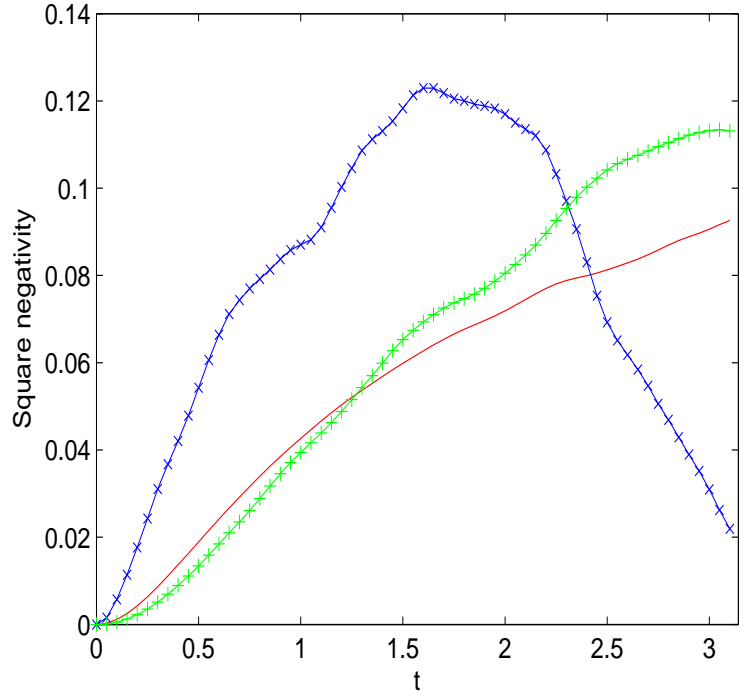,height=100mm,width=110mm}}    
\caption{(Color online) Comparison of `square-negativity' for classical, semi-quantum and  
semi-classical evolutions generated by $H=H_0^3/E^2$ for the time-interval $[0,\pi]$,  
and with $m=\omega=E=1$ and $\kappa=2,\mu=1/2,\alpha_0=0.5$.  
The solid line, the solid line with x points and the solid line with  
+ points represent the classical, semiquantum and semiclassical values respectively.
\label{negativity}}  
\end{figure}

The semiclassical eigenvalues remain very  
close to the classical eigenvalues over the whole time range. Since the  
behaviour of the individual eigenvalues does not seem to reflect 
the differences shown in the graphs of  
the first moment, this leads one to wonder if those differences  are reflected in  
some global property of the spectrum. We know that in each case, the evolved  
Groenewold operators have trace and square trace equal to 1, but we can  
look at the contribution to these quantities by the negative eigenvalues.  
Numerical experiments indicate that it is difficult to calculate the sum of  
the negative eigenvalues by using truncation techniques, so we instead  
concentrate on the sum of the squares of the negative eigenvalues.  
In Fig.(\ref{negativity}) we graph this `squared negativity' for the parameter values 
used in Fig.(\ref{sextic_alphaA})  

These results, however, merely confirm the conclusions drawn from
the eigenvalue curves 
and indicate  
that the spectrum of the semiclassical Groenewold operator resembles  
that of the classical Groenewold operator more closely than the spectrum of the 
semiquantum Groenewold operator does, for all but very small times. We 
conclude that the moments and the eigenvalue spectra provide quite different
information about these different approximate dynamics.

\section{Concluding remarks}
Our study indicates that semiquantum approximations to classical dynamics can provide 
new information about the interface between quantum and classical mechanics.
Semiquantum approximations obtained from the 
Hilbert space formulation of classical mechanics, and
semiclassical
ones obtained from the phase space formulation
of quantum mechanics, show significantly different behaviours for expectation values of observables
and eigenvalues of the (pseudo)density operator. 
For the simple systems considered here,
it appears that a first semiquantum approximation to quantum dynamics
is, for simple indicators like moments of coordinates and momenta,
closer to classical dynamics and further from quantum dynamics
than is a first semiclassical approximation to classical dynamics.  On the other hand,
as regards the spectrum of the density operator and Groenewold operator,  the first
semiclassical approximation behaves more like classical dynamics than the
first semiquantum approximation, which behaves more like quantum dynamics in this respect. 
It is not at all clear why this should be so, and it
is  desirable in future work to look at cases where higher approximations come into play to
see what happens then, as well as to examine the underlying theory more closely.

We have been able to consider only very simple one-dimensional systems here,
and there is obviously a need also 
to explore
systems with more degrees of freedom, especially nonintegrable ones.
The possible role of semiquantum mechanics in throwing new light on the interface between classical chaos and 
corresponding quantum dynamics
is especially interesting.

More generally, there are deep theoretical questions that arise 
about the mathematical relationship between
semiquantum and semiclassical approximations, associated with the structure
of the series expansions (\ref{class_evolution}) and (\ref{quant_evolution}). 
Another concerns the nature of the relationship between action 
principles and semiquantum approximations to
classical dynamics, whether formulated in phase space or Hilbert space.  
We hope to return to some of those questions.
\vskip1cm
\noindent
{\bf Acknowledgement:} We thank a referee for several useful comments and for bringing the papers of
Vercin \cite{Vercin00} and Voros \cite{Voros89} to our attention. 
This work was supported by Australian Research Council Grant DP0450778.

%\appendix    
    
\end{document}